\documentclass[twoside,a4paper]{article}
\usepackage{amsmath,amsfonts,amssymb,bm,mathrsfs}
\usepackage{color}
\usepackage{graphicx}
\graphicspath{{Figs/}}
\definecolor{MidRed}{rgb}{0.78,0,0}
\definecolor{MidGreen}{rgb}{0,0.65,0}
\definecolor{MidBlue}{rgb}{0,0,0.68}

\newcommand{\unit}[1]{\ensuremath{\mathrm{#1}}}
\newcommand{\ohm}{\ensuremath{\mathrm{\Omega}}}
\newcommand{\Celsius}{\ensuremath{\mathrm{^\circ C}}}
\newcommand{\req}[1]{(\ref{#1})}

\newcommand{\Memo}[1]{\qquad\textsf{\small(#1)}}

\def\rboth{\sffamily\tiny Y.\,Gruson, V.\,Giordano, U.\,L.\,Rohde, A.\,K.\,Poddar, E.\,Rubiola -- Cross-Spectrum PM Noise Measurement, Thermal Energy and Metamaterials} 
\author{Yannick Gruson%
\thanks{CNRS FEMTO-ST Institute, Dept.\ of Time and Frequency, Besancon, France.}, 
Vincent Giordano%
\thanks{CNRS FEMTO-ST Institute, Dept.\ of Time and Frequency, Besancon, France.},
\\[0.5ex]
Ulrich L. Rohde%
\thanks{Brandenburg University of Technology, Cottbus-Senftenberg, Germany, and Synergy Microwave Corp., McLean Blvd, Paterson, NJ, USA.},
Ajay K. Poddar%
\thanks{Synergy Microwave Corp., McLean Blvd, Paterson, NJ, USA.}, 
and Enrico Rubiola%
\thanks{CNRS FEMTO-ST Institute, Dept.\ of Time and Frequency, 26 Rue de l'Epitaphe, Besancon, France.   ER is the reference author, E-mail: rubiola@femto-st.fr., home page http://rubiola.org.}
}
\title{Cross-Spectrum PM Noise Measurements,\\Thermal Energy and Metamaterial Filters}

\begin{document}
\maketitle

\begin{abstract}
Virtually all commercial instruments for the measurement of the oscillator PM noise make use of the Cross Spectrum method (arXiv:1004.5539 [physics.ins-det], 2010).
High sensitivity is achieved by correlation and averaging on two equal channels which measure the same input, and reject the background of the instrument.
We show that a systematic error is always present if the \emph{thermal energy} of the input power splitter is not accounted for.  Such error can result in noise \emph{under estimation} up to a few dB in the lowest-noise quartz oscillators, and in an invalid measurement in the case of cryogenic oscillators.
As another alarming fact, the presence of \emph{metamaterial components} in the oscillator results in \emph{unpredictable behavior} and \emph{large errors}, even in well controlled experimental conditions.  We observed a spread of 40 dB in the phase noise spectra of an oscillator, just replacing the output filter.
\end{abstract}

\clearpage
\tableofcontents

\section{Introduction and State of the Art}\label{sec:intro}
\begin{figure}\centering
\includegraphics[scale=0.66]{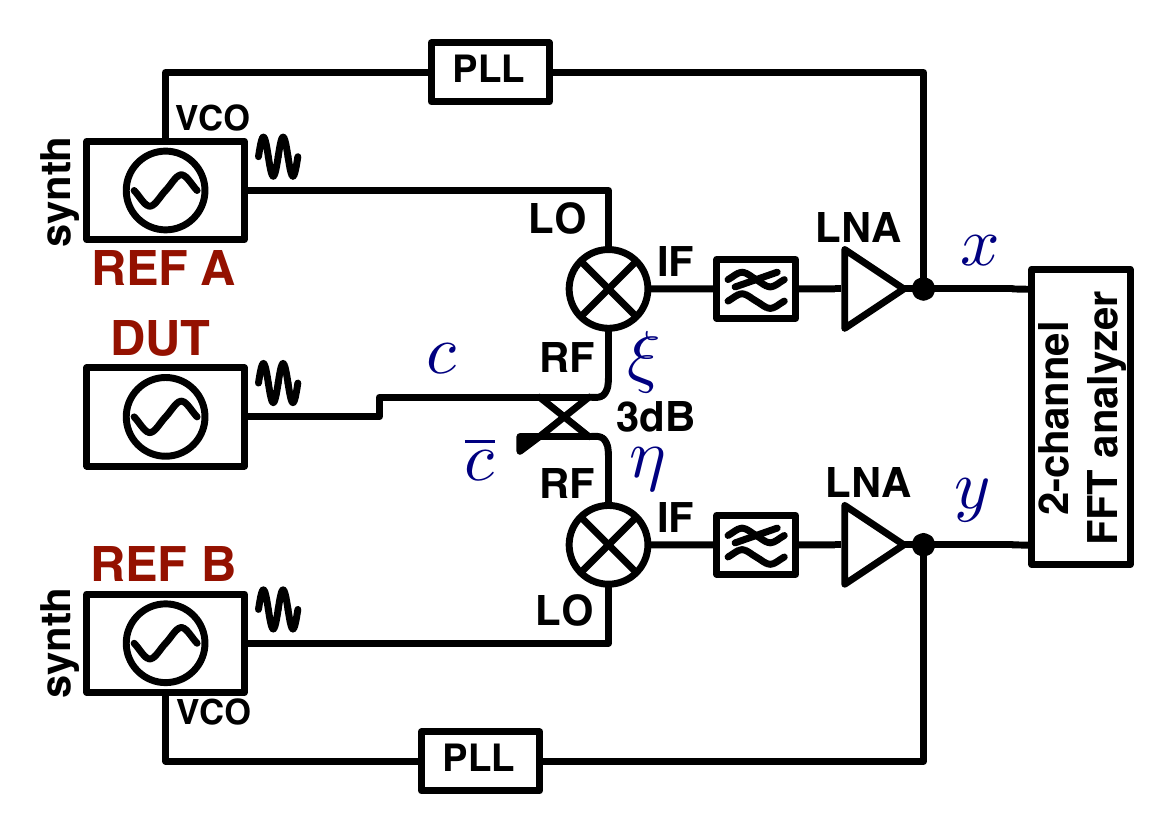}
\caption{Dual-channel phase noise measurement system.}
\label{fig:Nist-scheme}
\end{figure}

The dual-channel scheme shown in Fig.\,\ref{fig:Nist-scheme} is the standard method for the measurement of the oscillator phase noise, adopted by most manufacturers of instruments.  The main reason is that the single-channel background noise is averaged out, assuming that the two channels are statistically independent. 
This relaxes the requirement for reference oscillators and mixers having lower noise than that of the oscillator under test.
Modern digital electronics provides `killer' averaging power for cheap, compared to the cost of the RF section.

The method derives from early works in radio astronomy \cite{Hanbury-Brown-1952} and from the masurement of frequency fluctuations in H masers \cite{Vessot-1964}.  The first application to phase noise we could track came from Walls et.\ al.\ at NBS (former name of NIST) in 1976 \cite{Walls-1976}, using a dedicated fully-analog spectrum analyzer.
The cross spectrum become practical only later, when commercial FFT analyzers were available \cite{WWalls-1992}.
Since then, the development was mostly technical and commercial.  The method was left aside by the scientific community, and had been almost absent from the literature for a long time. 
Recently, Nelson et al.\ \cite{Nelson-2014} came up with simulations and a collection of spectra with artifacts and anomalies, pointing out the presence of a problem.  We tackled the problem at two workshops \cite{Paris-2014,Denver-2015}, yet without coming to a clear conclusion.

This article discusses two key facts, the role of thermal energy, and a weird effect of meta-material components.

In a two-channel system like Fig.\,\ref{fig:Nist-scheme}, the thermal noise of the dark port in the power splitter is anticorrelated.  This fact has been used extensively for decades in radiometry and Johnson thermometry \cite{Allred-1962,White-1996-Thermometry}.  In our earlier work \cite{Rubiola-2000-RSI,Rubiola-2002-RSI}, we proved that the cross-spectrum instrument cancels the thermal energy and measures the excess noise only, and we demonstrated a background of $-210$ dBc/Hz (white) and $-175$ dBc/Hz at 1 Hz offset (flicker).
However, at that time we focused on the $1/f$ noise of two-port components.  Regretfully, we did not realize that the same mechanism yields systematic errors in the measurement of oscillators.
The idea of anticorrelated noise from the input power splitter in the measurement of oscillators was suggested by Joe Gorin at the 2015 workshop \cite{Denver-2015}, and later analyzed by Hati \& al \cite{Hati-2016-xSpectrum}.  Working in parallel, we come to similar conclusions.  However, we rely on significantly different methods, and we provide the analytical theory.

Meta-materials have being long studied in optics and electromagnetics for various purposes, the most known of which is improving antennas and lenses (for example, \cite{Kock-1946}).  Such materials enable slow-wave propagation, negative permettivity $\epsilon$ and $\mu$, and even negative refraction index \cite{Shelby-2001}.  These possibilities are of obvious interest for resonators and filters \cite{Pond-2000}, and more recently for low-noise oscillators \cite{Poddar-2016-IFCS-Mobius,Rohde-2016-MJ1,Rohde-2016-MJ2,Rohde-2016-MJ3,Poddar-2014-habil}.
However, we observed experimentally that the use of a meta-material filter at the oscillator output results in erratic results, with a large spread of spectral values.  This fact is reported as a new warning to the community, with still no explanation.

\section{Dual-Channel Phase Noise Measurement}\label{sec:dual}
\begin{table}
\caption{Quantities and Notation}\label{tab:notation}
~\\
\begin{tabular}{ll}
$a$, $b$ & the random noise introduced by the two channels\\
$c$ & the random noise of the DUT\\
$d$ & a disturbing signal, random or deterministic\\
$k_d$ & the phase-to-voltage gain\\
$S$ & one-sided Power Spectral Density (PSD)\\
$\varphi$ & the DUT phase noise, $c=k_d\varphi$\\
$\varsigma_x$, $\varsigma_y$ & coefficients, either positive or negative\\
$\varsigma$ & a shorthand for $\varsigma_x\varsigma_y$\\
$\theta$ & $d$ converted to phase noise, i.e., $d=k_d\theta$\\
$\mathbb{E}\{\;\}$ & mathematical expectation operator\\
$\left<\;\right>_m$ & average on $m$ realizations\\
`hat' & estimator ($\hat{S}$ is an estimator of $S$)\\
\end{tabular}
\end{table}

\noindent Let us first introduce the following quantities and notation shown on Table~\ref{tab:notation}.

\subsection{Simplified Analysis}
We start with a generic dual-channel instrument, not limited to phase noise, and simpler than Fig.\,\ref{fig:Nist-scheme}.  Such instrument measures $c$ by correlating two equal and independent branches, whose outputs are 
\begin{align}
x&=c+a+\varsigma_xd &\leftrightarrow&& X&=C+A+\varsigma_xD\label{eqn:x}\\
y&=c+b+\varsigma_yd &\leftrightarrow&& Y&=C+B+\varsigma_yD\label{eqn:y}\,.
\end{align}
Here ``$\leftrightarrow$'' stands for the Fourier transform inverse-transform pair, time and frequency ($t$ and $f$) are implicit in the use of lowercase letters for the function of time, and of uppercase letters for the Fourier transform.  All the statistically independent noise goes in $a$ and $b$, and all the correlated noise goes in $c$ and $d$.  The correlated noise is either the DUT noise $c$, or the disturbing quantity $d$.
This formulation is therefore complete.
Of course, it may be necessary to split $d$ in separate processes, each governed by its own physical laws.
The two channels are nominally equal except for $\varsigma_x$ and $\varsigma_y$, which reflects the positive or negative correlation of $d$.

The cross PSD of a wide-sense stationary and ergodic signal is evaluated through the Fourier transform
\begin{align}
S_{yx}=\frac{2}{\mathcal{T}}\:YX^\ast
\Memo{one-sided cross PSD}.
\label{eqn:xpsd}
\end{align}
The factor `2' accounts for the power at negative frequencies, folded to positive frequencies, $\mathcal{T}$ is the acquisition time for each realization (we may let $\mathcal{T}\rightarrow\infty$ in theoretical issues), and 
the superscript `$\ast$' means  complex conjugate.  

Averaging out the single-channel noise, the cross PSD is  
\begin{align}
S_{yx}=\:\frac{2}{\mathcal{T}}\left[|C|^2 + \varsigma\left|D\right|^2\right] = S_c+\varsigma S_d\,.
\label{eqn:Syx}
\end{align}
If $d$ is not accounted for, the term $\varsigma S_d$ turns into a systematic error which can be either positive or negative, i.e., over-estimation or under-estimation of the DUT noise.

\subsection{Spectral Estimation}
However simple, the proof of \req{eqn:Syx} provides insight.
Using the superscript `prime' and `second' for real and imaginary part, as in $A=A'+iA''$, and expanding $YX^\ast$, we find
\begin{align}
\Re\{YX^\ast\}
&=|C|^2+\varsigma|D|^2\nonumber\\
&+A'B'+A''B''+A'C'+A''C''+B'C'+B''C''\nonumber\\
&+\varsigma_y(A'D'+A''D'')+\varsigma_x(B'D'+B''D'')\nonumber\\
&+(\varsigma_y+\varsigma_x)(C'D'+C''D'')
\label{eqn:Re}\\[1ex]
\Im\{YX^\ast\}
&=A'B''-A''B'+A'C''-A''C'-B'C''+B''C'\nonumber\\
&+\varsigma_y(A'D''-A''D')-\varsigma_x(B'D''-B''D')\nonumber\\
&+(\varsigma_y-\varsigma_x)(C'D''-C''D')\,.
\label{eqn:Im}
\end{align}
It is worth mentioning that the cross terms $C'C''$ and $D'D''$ cancel.  This is interesting because the real and the imaginary part of a parametric noise process can be correlated.

Taking the mathematical expectation, we get
\begin{align}
\mathbb{E}\bigl\{YX^\ast\bigr\} 
= \mathbb{E}\bigl\{|C|^2\bigr\} + \varsigma\mathbb{E}\bigl\{|D|^2\bigr\} + i0\,.
\label{eqn:Expectation}
\end{align}
The term `$i0$' emphasizes the fact that all the useful signal goes in $\Re\{YX^\ast\}$, thus in $\Re\{S_{yx}\}$.

Actual measurements rely on an estimator.  After \req{eqn:Expectation},
\begin{align}
\hat{S}_{yx}=\smash{\bigl<\Re\{S_{yx}\}\bigr>}_m
\Memo{best estimator}
\label{eqn:estimator}
\end{align}
is an obvious choice because this estimator is unbiased
\begin{align*}
\mathbb{E}\{\hat{S}_{yx}\}
=\mathbb{E}\{\smash{\bigl<\Re\{S_{yx}\}\bigr>}_m\}
=\mathbb{E}\left\{S_{yx}\right\}\,.
\end{align*}
For finite $m$, it can be shown that \cite{Rubiola-2010-arXiv-xSpectrum}
\begin{align*}
\hat{S}_{yx} = \mathbb{E}\{S_{yx}\}+\mathcal{O}(1/\sqrt{m})\,.
\end{align*}
The notation $\mathcal{O}(\,)$ means `order of,' as used with truncated series and polynomials.  Here, $\mathcal{O}(1/\sqrt{m})$ represents the single-channel noise, which is rejected with $1/\sqrt{m}$ law.  As a consequence, an additional 5 dB rejection costs a factor of 10 in the averaging size, thus in measurement time. 

Most instruments use the estimator 
\begin{align}
\hat{S}_{yx}=\bigl|\smash{\bigl<S_{yx}\bigr>}_m\bigr|
\Memo{practical estimator}.
\label{eqn:biased-estimator}
\end{align}
This choice fits most practical needs, and is always suitable to logarithmic (dB) display.  

Besides bias, which is obvious in \req{eqn:biased-estimator}, we have two reasons to prefer \req{eqn:estimator} to \req{eqn:biased-estimator}.   
First, \req{eqn:biased-estimator} takes larger $m$ to achieve the same noise rejection, thus longer measurement time (a factor of four applies in the case of white noise).  This happens because the background noise is split between $\Re\{S_{yx}\}$ and $\Im\{S_{yx}\}$.
Second, the notches pointed out by Nelson et al.\ \cite{Nelson-2014} are a good argument in favor of \req{eqn:estimator} versus \req{eqn:biased-estimator}.
Such notches may hide a case where $\varsigma<0$ and $S_d$ approaches or exceeds $S_c$, which is a total nonsense.

\subsection{Application to Phase Noise}
For the measurement of phase noise (Fig.\,\ref{fig:Nist-scheme}) we replace
\begin{align*}
c &\rightarrow k_d\varphi & S_c &\rightarrow k^2_dS_\varphi\\
d &\rightarrow k_d\theta  & S_d &\rightarrow k^2_dS_\theta\,,
\end{align*}
where $\varphi$ is the DUT phase noise, and $\theta$ is the correlated noise $d$ converted into phase noise.  Hence 
\begin{align}
S_\varphi
=\frac{1}{k_d^2}\Bigl[S_{yx}-\varsigma S_d\Bigr]
=\frac{1}{k_d^2}S_{yx}-\varsigma S_\theta\,.
\label{eqn:Sphi}
\end{align}
The instruments are generally not aware of $d$.  Accordingly, the readout equation $S_\varphi=\frac{1}{k_d^2}S_{yx}$ introduces a systematic error $\Delta S_\varphi=\varsigma S_\theta$.

\section{Thermal Energy in the Input Power Splitter}\label{sec:splitter}
We introduce the thermal noise of the input power splitter, and we analyze its consequences on the measurement of $S_\varphi$.
The simplest way to understand the problem is to focus on the thermal noise associated to the RF signals $\xi$ and $\eta$ in Fig.\,\ref{fig:Nist-scheme} (see also Fig.~\ref{fig:Dir-cpl} and \ref{fig:Res-cpl}, discussed later).  The PSD of the available voltage is equal to $kTR_0$, where $k=1.38{\times}10^{-23}$ is the Boltzmann constant, $T$ is the equivalent temperature, and $R_0$ the characteristic impedance.
Engineers often use the `thermal emf in 1 Hz bandwidth,' denoted with $e_n$ and ruled by $\mathbb{E}\{e^2_n\}=4kTR$ for thermal noise; the expectation $\mathbb{E}\{\:\}$ is usually implied.
For reference, $e_n$ is of 0.9 \unit{nV/\sqrt{Hz}} for a 50 \ohm\  resistor at room temperature, and  1--5 \unit{nV/\sqrt{Hz}} for a good operational amplifier, depending on the requirements on the noise current $i_n$.

Following our approach, the phase detector is seen as a \emph{receiver} described in terms of \emph{back scatter} temperature $T^\star_R$.
The reason for separating $T^\star_R$ from the `regular' noise temperature $T_R$ is that the noise radiated back from the output generates crosstalk, and in turn it contributes to the background noise.  By contrast, $T_R$ includes other noise processes which are not back-scattered, and averaged out in the dual-channel scheme.
The receiver can be a double balanced mixer as in Fig.\,\ref{fig:Nist-scheme}, or the more sensitive detector we used in \cite{Rubiola-2000-RSI,Rubiola-2002-RSI}.

The problem is therefore to estimate the additive noise $S_c=kT_CR_0$ of signal $c$ from $S_{yx}$, and then to calculate $S_\varphi$ using
\begin{gather}
S_\varphi(f)=\frac{kT_C}{P_0}=\frac{S_c}{R_0P_0}\,,
\label{eqn:b0}
\end{gather}
where $P_0$ is the carrier power.

\subsection{Loss-Free Power Splitter}\label{sec:Dir-cpl}
\begin{figure}\centering
\includegraphics[scale=0.6]{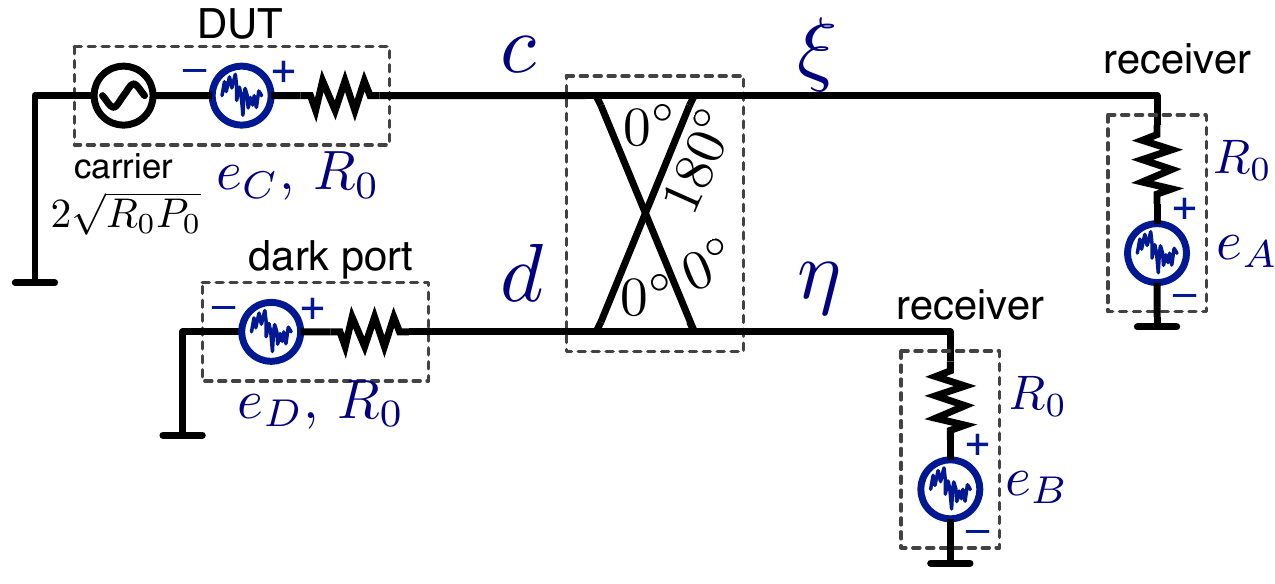}
\caption{Directional coupler used as the 3-dB power splitter, and noise emfs.}
\label{fig:Dir-cpl}
\end{figure}
The 4-port directional coupler terminated at one input (dark port) is by far the preferred power splitter (Fig.\,\ref{fig:Dir-cpl}).
Dropping the carrier, the signals at the output of the coupler are
\begin{align*}
x &= \frac{1}{2\sqrt{2}}\Bigl[e_C-e_D\Bigr] + \frac{1}{2}e_A\\[1ex]
y &= \frac{1}{2\sqrt{2}}\Bigl[e_C+e_D\Bigr] + \frac{1}{2}e_B\,.
\end{align*}
The equivalent temperature is $T_C=P_0S_\varphi/k$ for the oscillator's noise floor, and $T_D$ for the dark port.
A simple calculation gives
\begin{align}
\mathbb{E}\{S_{\eta\xi}\}
&= \frac{1}{8}\Bigl[\mathbb{E}\left\{e^2_C\right\}-\mathbb{E}\left\{e^2_D\right\}\Bigr]\nonumber\\[1ex]
&= \frac{1}{2}k\Bigl[T_C-T_D\Bigr]R_0\,.
\label{eqn:Syx-Cpl}
\end{align}

Equation \req{eqn:Syx-Cpl} is the physical principle of the correlation radiometer \cite{Allred-1962}, and also used in Johnson thermometry \cite{White-1996-Thermometry}.  In this case, $T_D$ is the reference temperature, chiefly the triple point of \unit{H_2O}, and the instrument measures the difference $\Delta T=T_C-T_D$.  The preferred estimator is 
\begin{gather}
\widehat{\Delta T}=\frac{2}{kR_0}\bigl<\Re\bigl\{S_{\xi\eta}\bigr\}\bigr>_m
\Memo{thermometer}\,.
\label{eqn:Delta-T}
\end{gather}

A rigorous evaluation of $S_\varphi$ can be done combining \req{eqn:b0} and \req{eqn:Syx-Cpl}.  There results the estimator
\begin{align}
\hat{S}_\varphi 
&= \frac{2}{R_0P_0}\:\left<\Re\{S_{\eta\xi}\}\right>_m + \frac{kT_D}{P_0}
\Memo{unbiased},
\label{eqn:Sphi-cpl-right}
\intertext{or equivalently}
\hat{S}_\varphi 
&= \frac{1}{k^2_d}\:\left<\Re\{S_{yx}\}\right>_m + \frac{kT_D}{P_0}
\label{eqn:Sphi-cpl-right-out}
\end{align}
when observed from the output.

By contrast, using the estimator $\hat{S}_\varphi\left<\Re\{S_{yx}\}\right>_m/k^2_d$ we discard the thermal energy of the dark port.
This estimator suffers from a bias $\Delta S_\varphi=\hat{S}_\varphi-\mathbb{E}\{S_\varphi\}$ given by
\begin{align}
\Delta S_\varphi 
&= - \frac{kT_D}{P_0}
\Memo{bias}.
\label{eqn:Sphi-cpl-error}
\end{align}
This is a systematic \emph{under-estimation} of the DUT noise.

\subsection{Resistive Power Splitter}\label{sec:Res-cpl}
\begin{figure}\centering
\includegraphics[scale=0.6]{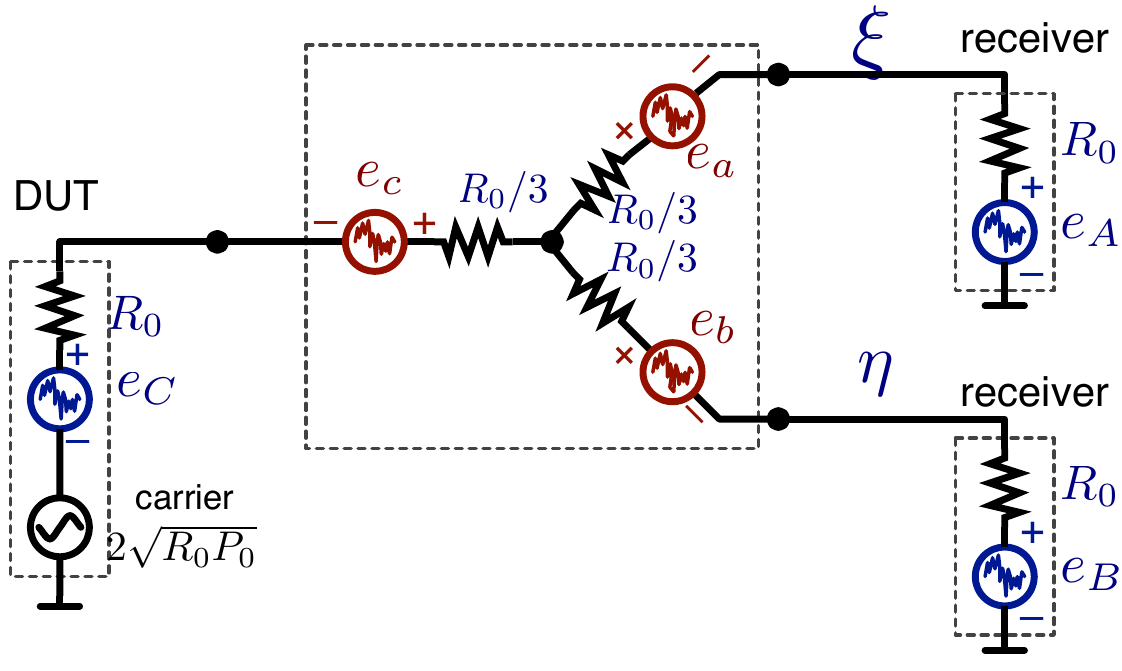}
\caption{Resistive (Y) 6-dB power splitter and noise emfs.}
\label{fig:Res-cpl}
\end{figure}
The Y resistive network (Fig.\,\ref{fig:Res-cpl}) is sometimes used instead as a power splitter.  To our knowledge, the Keysight E5500 series \cite{Keysight:E5500} is the one and only case where a commercial instrument uses the Y coupler.
Dropping the carrier, the random signals at the splitter output are 
\begin{align}
\xi&=\tfrac{1}{2}e_A+\tfrac{1}{4}e_B+\tfrac{1}{4}e_C-\tfrac{1}{2}e_a+\tfrac{1}{4}e_b+\tfrac{1}{4}e_c\\[1ex]
\eta&=\tfrac{1}{4}e_A+\tfrac{1}{2}e_B+\tfrac{1}{4}e_C+\tfrac{1}{4}e_a-\tfrac{1}{2}e_b+\tfrac{1}{4}e_c\,.
\end{align}
For the evaluation of the thermal emfs,
the DUT temperature is $T_C$,
the receiver backscatter temperature is $T^\star_R$, and
the temperature of the splitter's internal resistors is $T_S$.
The cross PSD at the splitter output is 
\begin{align}
\mathbb{E}\{S_{\xi\eta}\}
&=k\left[\frac{1}{4}T_C-\frac{1}{4}T_S+T^\star_R\right]R_0\,.
\label{eqn:Syx-Res}
\end{align}
The rigorous evaluation of $S_\varphi$ results from combining $S_\varphi=kT_C/P_0$ \req{eqn:b0} with \req{eqn:Syx-Res}
\begin{align}
\hat{S}_\varphi 
&= \frac{4}{R_0P_0}\:\hat{S}_{\eta\xi} + \frac{k(T_S-4T^\star_R)}{P_0}
\label{eqn:Sphi-res-right}
\Memo{unbiased},
\intertext{or equivalently}
\hat{S}_\varphi 
&= \frac{1}{k_d^2}\:\hat{S}_{yx} + \frac{k(T_S-4T^\star_R)}{P_0}
\label{eqn:Sphi-res-right2}
\end{align}
as seen at the output.
Neglecting the splitter's thermal energy, the estimator \req{eqn:Sphi-res-right2} becomes $\hat{S}_\varphi=\hat{S}_{yx}/k^2_d$.  There results a bias bias error given by
\begin{align}
\Delta S_\varphi 
&= - \frac{k(T_S-4T^\star_R)}{P_0}
\Memo{bias}.
\label{eqn:Sphi-Y-error}
\end{align}
Unlike the directional coupler, there is no general rule to assess whether $\Delta S_\varphi$ is positive or negative.

\section{Low-Noise Oscillators}\label{sec:ln-oscillators}
\begin{figure}\centering
\includegraphics[scale=0.64]{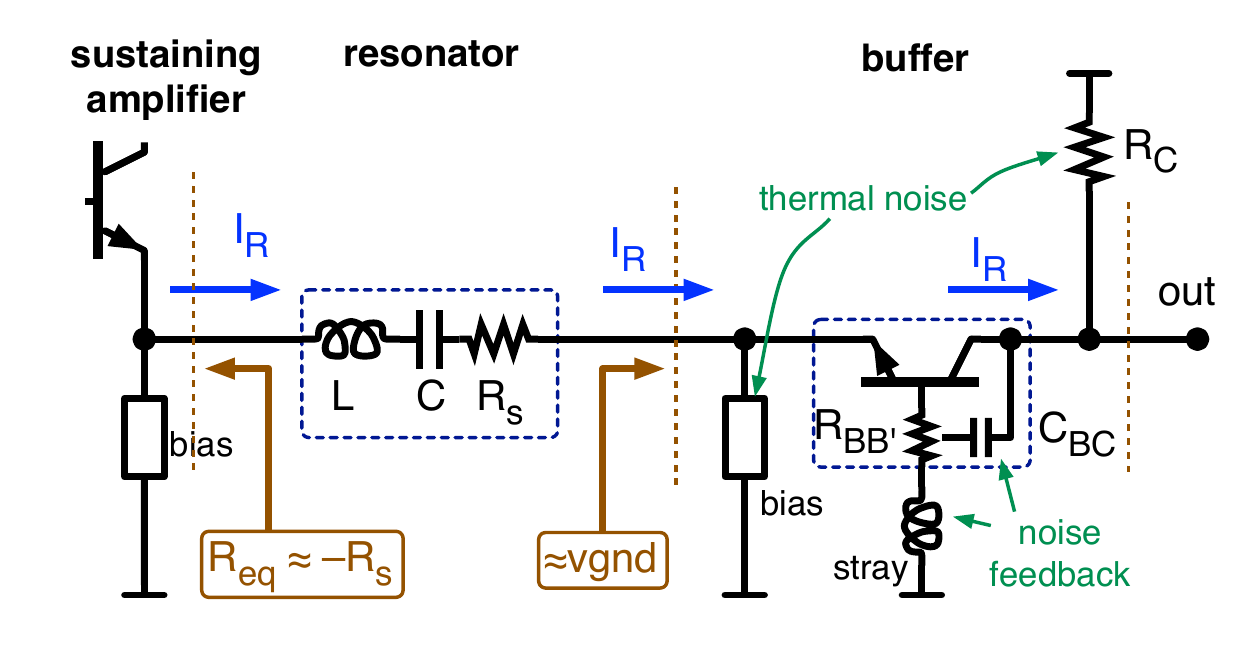}
\caption{Principle of ultra-low noise floor oscillators.}
\label{fig:Oscillator}
\end{figure}

Understanding the interplay of noise sources which give rise to  errors is a part of the message we address to the reader.
The cryogenic oscillator is rather a trivial case, to the extent that it is conceptually simple to achieve low $T_C$ by cooling the entire oscillator, and in turn to experience errors or invalid $S_{yx}<0$.

A more subtle case is a class of oscillators which feature low equivalent temperature, close to the physical temperature, despite of the buffer's power gain introduced between the oscillator internal loop and the output.  This is achieved by inserting a bandpass filter between the oscillator loop and the buffer \cite{Rohde-1978}, and with a special design of the buffer. 
A simplified and more efficient version (Fig.\,\ref{fig:Oscillator}) uses the same resonator as the reference resonator and as the output filter.  This design solves the issue of harmonic distortion and circumvents the white phase noise of the oscillator at once \cite{Rohde-1975-ED} (See also \cite[p.~264]{Rohde:Synthesizers} for the electrical diagram of a complete oscillator).  
It goes without saying that VHF TCXOs, often at 100 MHz or 125 MHz, are made in this way when they are intended for special applications where the lowest white PM noise is the most desired feature.

The lowest-noise design relies on the following two ideas.  

The first is that the filter is strongly mismatched in the stop-band ($\Gamma{\,=\,}{\pm1}$ for $|\nu-\nu_0|>\nu_0/2Q$).  So, in the stop-band the noise is \emph{not coupled} to the buffer.  This holds for the noise of the sustaining amplifier, and also for the thermal energy associated to the dissipation inherent in the filter.
The consequence is that, beyond the Leeson frequency $f_L=\nu_0/2Q$, the white phase noise is limited only by the buffer noise divided by the carrier power.  
Some people are surprised to learn that the thermal energy associated to the motional resistance of the piezoelectric quartz does not contribute.

The second idea is the use of a grounded-base configuration for the buffer, which has very low input impedance. 
The transistor noise is chiefly described by the Giacoletto or Gummel-Poon models \cite{Giacoletto-1969,Gummel-Poon-1970} in the presence of internally generated shot and thermal noise \cite{Niu-2001,Voinigescu-1997} (see also \cite[Sec.~7.6]{Vasilescu-2005} for a useful discussion on noise models in the BJT).  In actual RF circuits, a positive feedback increases the noise.  This  is due to the interplay between the base-collector capacitance, the bond inductance and other parameters.  The overall effect is that the noise figure increases at higher carrier frequency, and nonetheless it may stay within 1 dB or less.
Finally, keeping the base-to-collector gain smaller than one, say 0.5, the transistor noise is attenuated rather than amplified.

As an example, a 120 MHz OCXO%
\footnote{The identity of this oscillator is not disclosed.
The reason is to prevent polemics which would inevitably bring us far from the concepts we want to illustrate.  However, search engines find a few similar  80--125 MHz oscillators with extremely low noise floor, similar to our example.} 
has  output power $P_0=20$ mW (13 dBm) and white PM noise $S_\varphi=5{\times}10^{-19}$ $\mathrm{rad^2/Hz}$ ($\mathscr{L}=-186$ dBc/Hz).  Using $S_\varphi=kT_C/P_0$, we get $kT_C=10^{-20}$ W/Hz ($-170$ dBm/Hz), thus $T_C=724$ K\@.   If the measure is biased by the thermal energy as explained in Section~\ref{sec:Dir-cpl}, the correct value is  $T_C=(T_C)_\text{readout}+T_\text{room}=1024$ K, hence $S_\varphi=7.1{\times}10^{-19}$ ($-181.5$ \unit{dBrad^2/Hz}), and the bias error is of $-1.5$ dB\@.

\section{Experiments Related to the Thermal Energy}\label{sec:thermal}
\begin{figure}\centering
\includegraphics[scale=0.64]{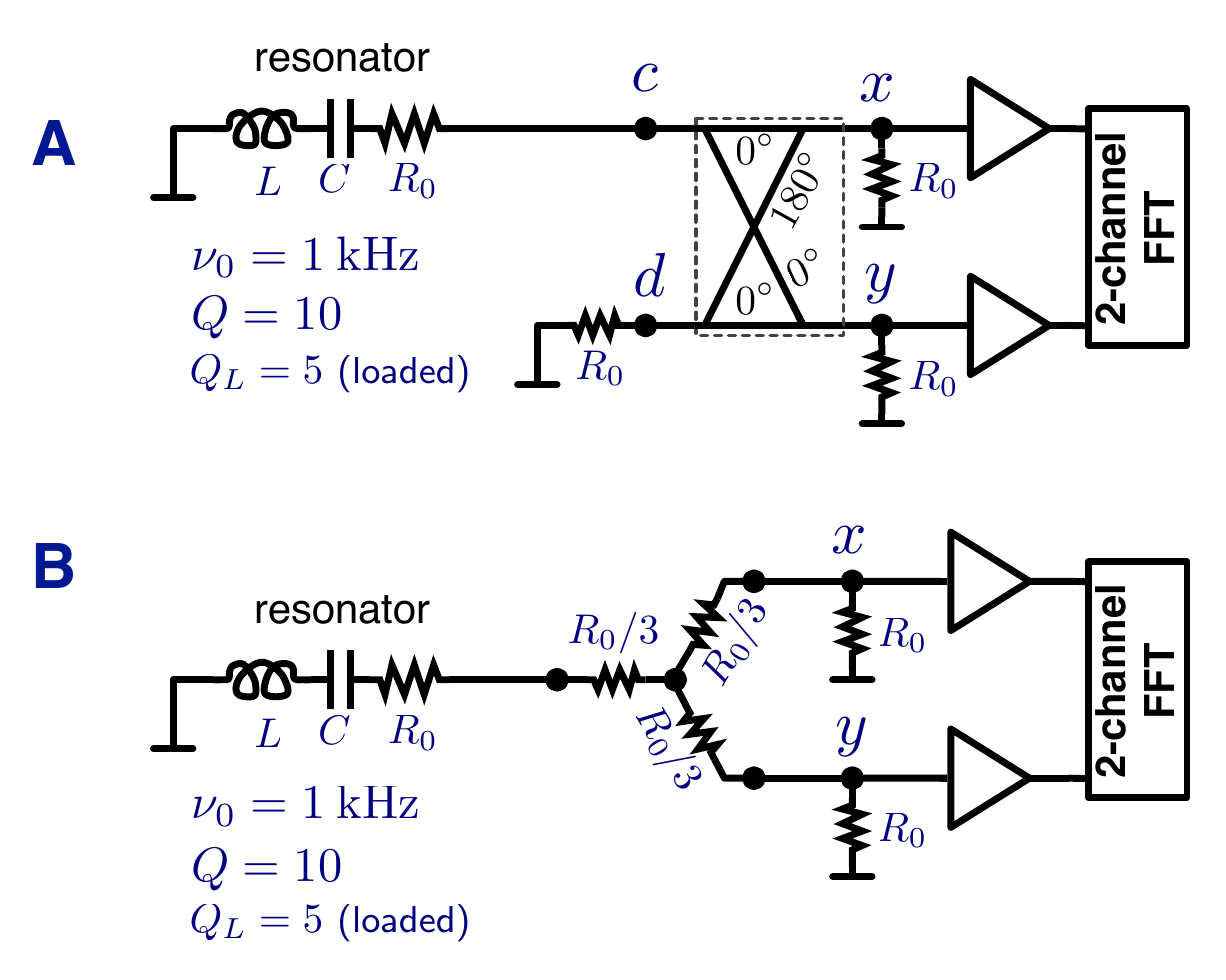}
\caption{Experimental configurations A and B.}
\label{fig:Experiments}
\end{figure}

Given the additive nature of the white PM noise, the presence of a carrier signal is not necessary, and we can work with the RF noise trusting $S_\varphi(f)=kT/P$.
Our experiments (Fig~\ref{fig:Experiments}) are appropriate to the internal configuration of the low-noise oscillators, where a resonator is used as the output filter \cite{Rohde-1978,Driscoll-1995-FCS}.  

Having said that, we decided to work on mockup where the frequency is scaled down by a factor of $10^5$, i.e., 1 kHz instead of 100 MHz.  The reason for this choice is that we have full control on both $T_R$ ant $T^*_R$ of the receiver.  Using a JFET operational amplifiers (AD\,743) and low $R_0$ (${<}10$ k\ohm) at the receiver front end, the input noise current $i_n$ is negligible.  Thus the backscatter radiation is the thermal energy of the input $R_0$, hence $T^*_R=T\approx300$ K, the physical temperature.  Adding the AD\,743's $e_n$ contribution, $T_R$ is of the order of 1500 K\@.  The amplifier gain is 120 (41.6 dB).
The FFT analyzer is a Hewlett Packard 5362A\@.  Yet, the same results are expected with any similar and more modern instrument.

\begin{figure}\centering
\includegraphics[scale=0.24]{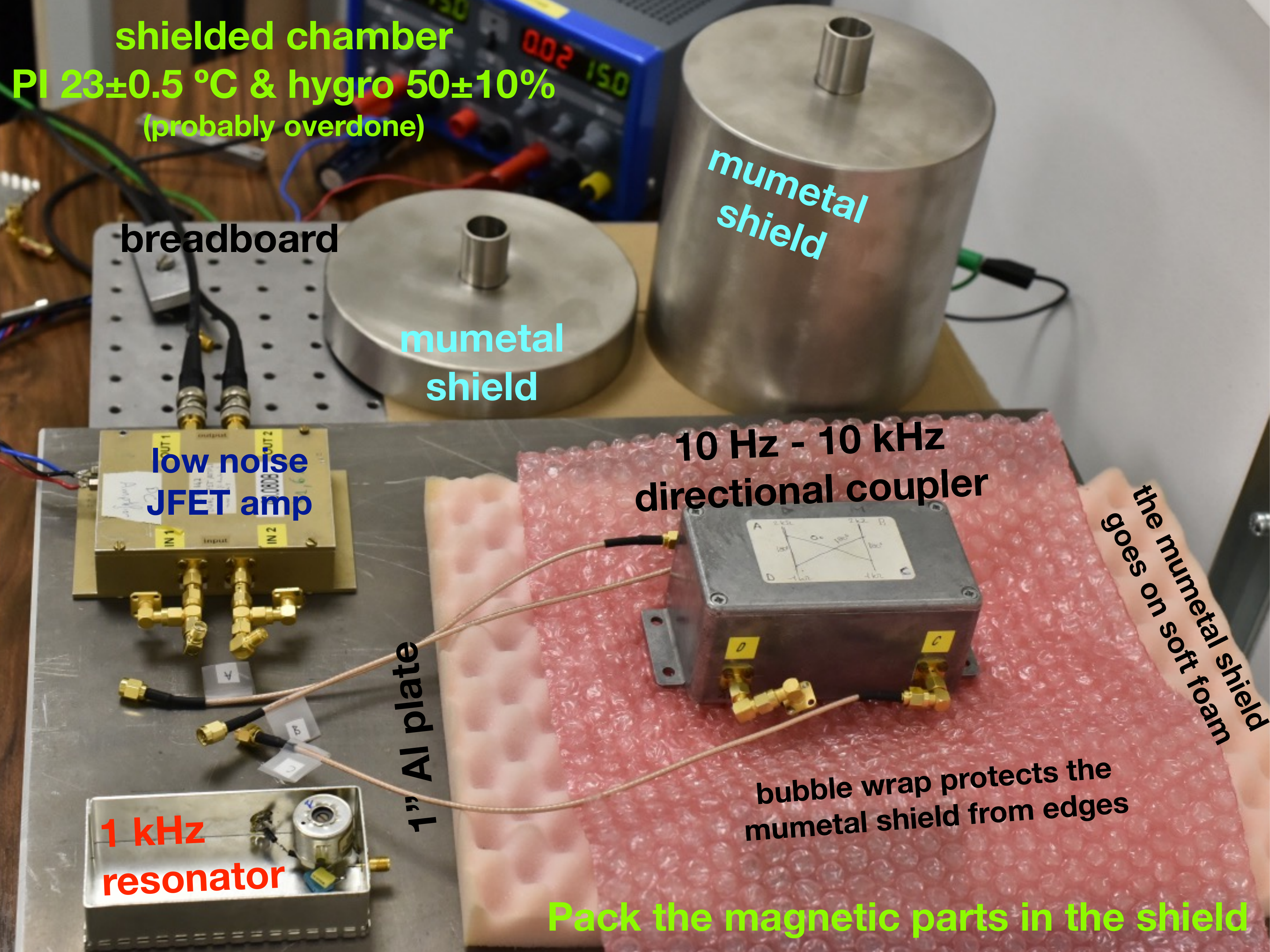}
\caption{Key parts of the experiments.}
\label{fig:Photo}
\end{figure}
These experiments are done at the FEMTO-ST Institute in a shielded chamber with proportional-integral control of temperature at $23\pm0.5$ \Celsius\ and humidity at $50\%\pm10\%$.  The critical circuits are further shielded in a mumetal enclosure borrowed from an atomic-clock experiment.  Putting the experiment on a 1'' Al plate improves short-term temperature stability, and also increases shielding thanks to eddy currents.  While temperature and humidity control is probably overdone, shielding is essential.  Figure~\ref{fig:Photo} shows the main components of the experiment, in their environment.

\subsection{First Experiment}
\begin{figure}\centering
\includegraphics[scale=0.55]{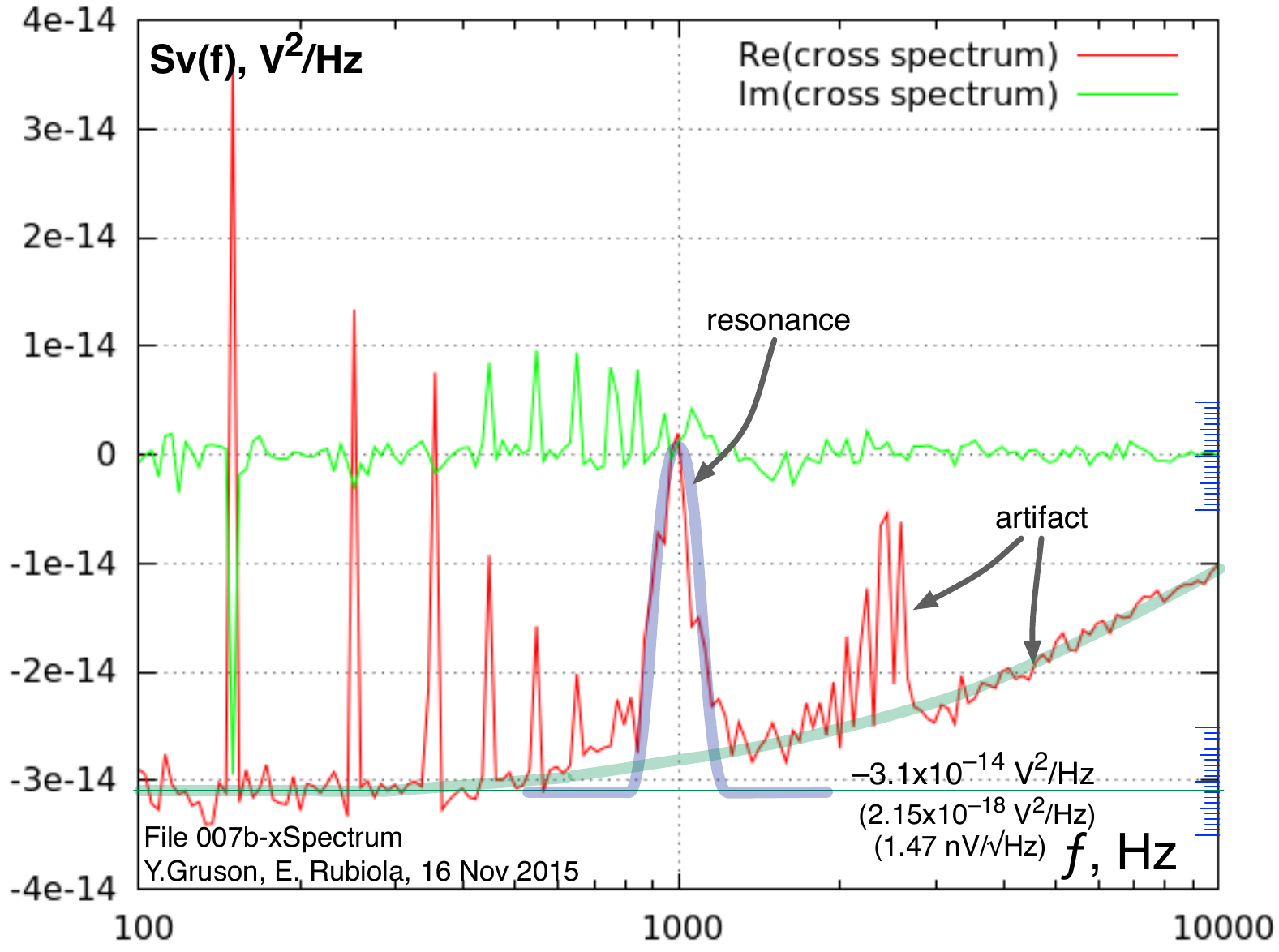}
\caption{Cross spectrum measured with the configuration of Fig.\,\ref{fig:Experiments}\,A.}
\label{fig:007b-xSpectrum}
\end{figure}

\begin{figure}\centering
\includegraphics[scale=0.55]{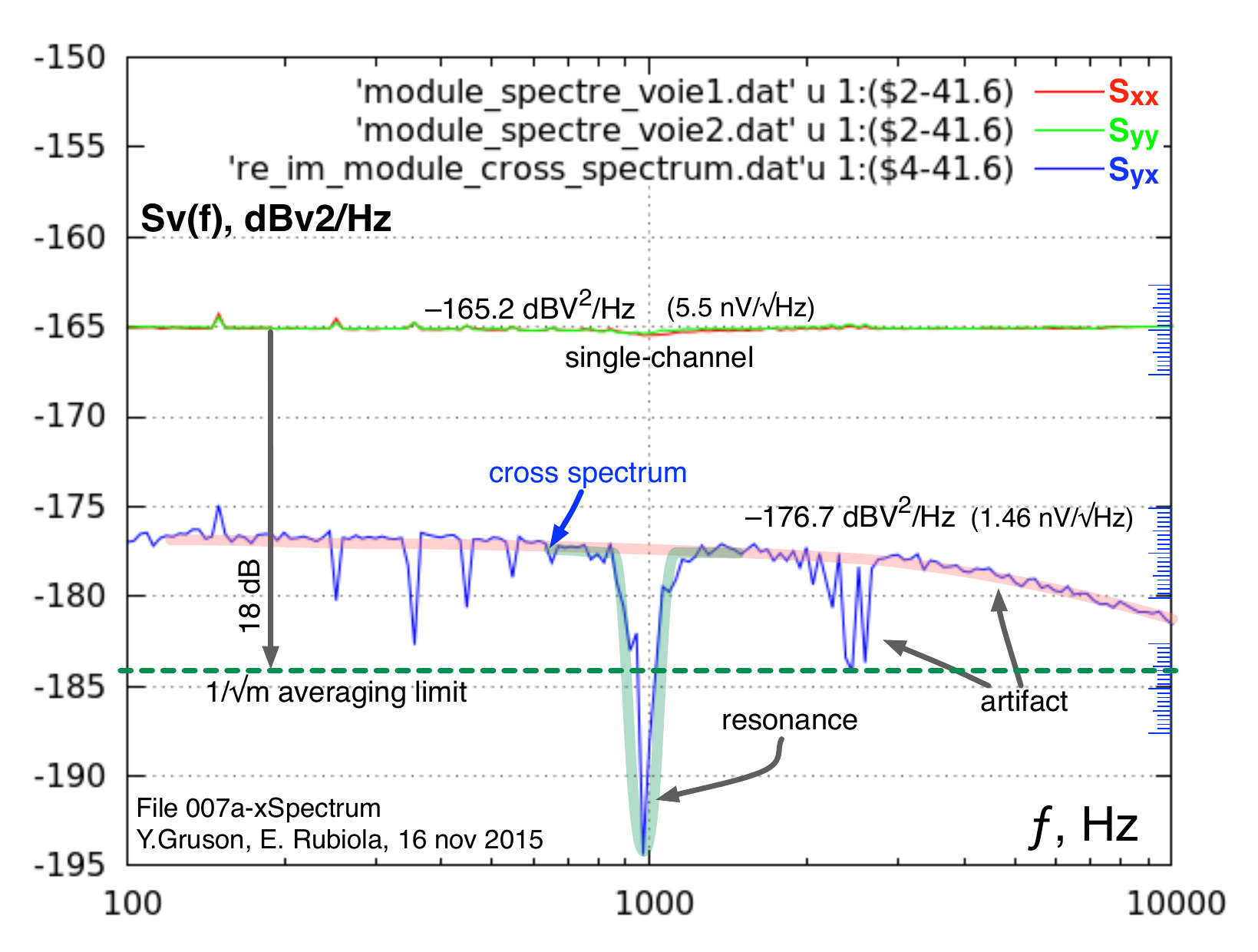}
\caption{Same run as in Fig.\,\ref{fig:007b-xSpectrum}, but $|S_{yx}|$ is shown on a log scale, together with the single-channel noise.}
\label{fig:007a-xSpectrum}
\end{figure}

In the first experiment (Fig.\,\ref{fig:Experiments}\,A) we use a custom directional coupler based on traditional transformers with laminated silicon steel core.  Re-using transformers from surplus lock-in amplifiers, we ended up with a trivial $1\,{:}\,\sqrt{2}$ voltage ratio.  Hence $R_0$ is 300 \ohm\ on the left-hand side and of 600 \ohm\ on the right-hand side.
The resonator is implemented with a 470 mH pot-type ferrite inductor and a mylar capacitor, resonating at 1 kHz with $Q=5$ ($Q=10$ with no load), and $R_0=300$ \ohm.
The spectrum, shown on Fig.\,\ref{fig:007b-xSpectrum}-\ref{fig:007a-xSpectrum}, is obtained averaging on $m\approx4000$ spectra.  The rejection of the single-channel noise, $1/\sqrt{m}$ is 18 dB.

Let us start from Fig.\,\ref{fig:007b-xSpectrum}, which shows $\Re\{S_{yx}\}$ and $\Im\{S_{yx}\}$.
At resonance, the coupler inputs receive the thermal noise from two equal resistors $R_0$ at the same temperature.  In this conditions, the peak of $\Re\{S_{yx}\}$ is close to zero, as predicted by Eq.~\req{eqn:Syx-Cpl}.  
Off resonance, the resonator is open circuit ($\Gamma=1$), hence its thermal noise does not appear in the signals $x$ and $y$.  By contrast, the signal $d$ appears in $x$ and $y$ with opposite signs, as seen from the $0^\circ$ and $180^\circ$ marks on Fig.\,\ref{fig:Experiments}A.  In this conditions, we measure $\Re\{S_{yx}\}=-2.2{\times}10^{-18}$ \unit{V^2/Hz}, in fairly good agreement with the expected value of $-2.5{\times}10^{-18}$  \unit{V^2/Hz}.  The latter is predicted by Eq.~\req{eqn:Syx-Cpl} with $R_0=600$ \ohm, at the right-hand side of the coupler.  
It is worth mentioning that $\Im\{S_{yx}\}$ is close to zero at all frequencies, as expected.

Figure\,\ref{fig:007a-xSpectrum} shows the same experimental data,
yet plotted as the absolute value.  The single-channel background noise is too high for the resonance to be visible on $|S_{xx}|$ and $|S_{yy}|$.  These plots overlap, as expected from symmetry.  
The resonance is seen as a dip at $f=1$ kHz on $|S_{yx}|$, and also some artifacts show up as dips.
Off resonance, $|S_{yx}|$ is 12 dB lower than the background noise.  

In summary, thermal noise of the dark port cancels the thermal noise at the input.  Since this `cancellation' still has effect outside the resonator bandwidth, where there is no noise to cancel, $S_{yx}$ is negative.  
This behavior is transposed to the measurement of oscillator by normalizing on the carrier power.  In the case of challenging low noise oscillators (Sec.\,\ref{sec:ln-oscillators}), the white PM region of $S_\varphi(f)$ may suffer from serious under-estimation, or even become negative.

\subsection{Second Experiment}
\begin{figure}\centering
\includegraphics[scale=0.55]{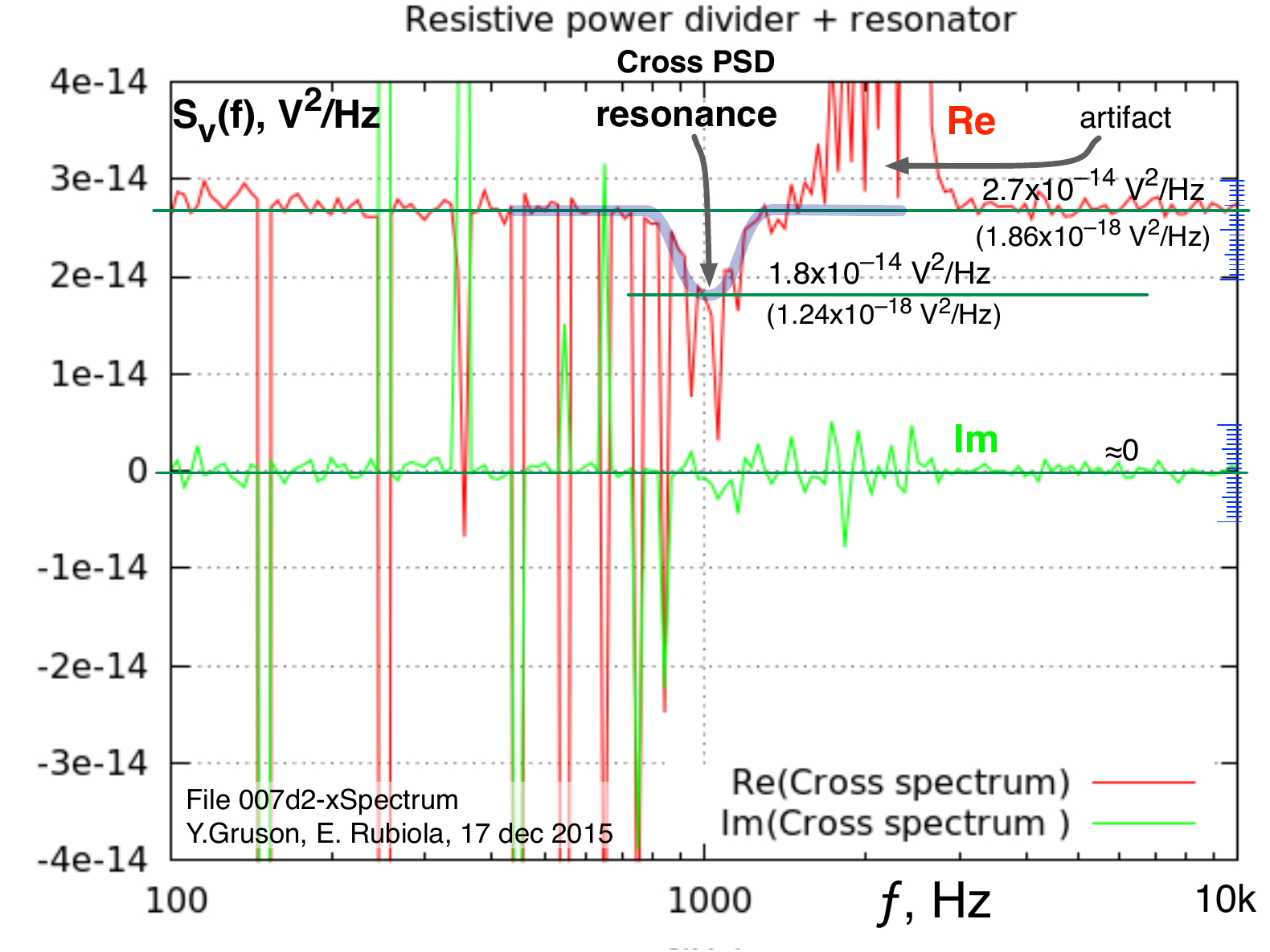}
\caption{Cross spectrum measured with the configuration of Fig.\,\ref{fig:Experiments}\,B.}
\label{fig:007d2-xSpectrum}
\end{figure}

The second experiment (Fig.\,\ref{fig:Experiments}\,B) uses a Y power splitter implemented with $R_0/3$ (100 \ohm) metal-film resistors, while the resonator is the same as in the first experiment.

At the resonance, \req{eqn:Syx-Res} predicts that $S_{yx}=kTR_0$ when the whole system is at the physical temperature $T$.  The value of $1.25{\times}10^{-18}$, seen on Fig.\,\ref{fig:007d2-xSpectrum}, is in a close agreement to \req{eqn:Syx-Res} with $T=300$ K and $R_0=300$ \ohm.

Off resonance, $e_c$ turns into open circuit, and the system changes performance.  From Fig.\,\ref{fig:Res-cpl}, we get 
\begin{align*}
S_{yx}
&=k\left[\frac{15}{16}T^*_A+\frac{15}{16}T^*_B-\frac{3}{8}T_S\right]R_0,
\intertext{and}
S_{yx}&=k\left[\frac{15}{8}T^*_R-\frac{3}{8}T_S\right]R_0\,
\end{align*}
assuming that back-scatter temperature of the two receivers is the same and equal to $T^*_R$, and that the power splitter is at the temperature $T_S$.  Finally, when the whole system is at the same temperature $T$, it holds that
\begin{align*}
S_{yx}=\frac{3}{2}kTR_0\,.
\end{align*}
The value observed on Fig.\,\ref{fig:007d2-xSpectrum} is $S_{yx}=1.86{\times}10^{-18}$ \unit{V^2/Hz}, and of course $\Im\{S_{yx}\}$ is close to zero at all frequencies.

\section{Experiments Related to Metamaterial Filters}\label{sec:metamat}
Homogeneous electrical resonators use `regular' materials, having positive refraction index $n{\,=\,}\sqrt{\mu\epsilon}$, where $\mu$ and $\epsilon$ are the magnetic permeability and the electric permittivity of the medium ($\mu_0=4\pi10^{-7}$ H/m and $\epsilon_0=1/\mu_0c^2\simeq8.85{\times}10^{-12}$ F/m for vacuum).
Their noise is usually modeled  using current and voltage sources, whose average values are obtained from the fluctuation dissipation theorem based on the normalized impedance \cite{Callen-Welton-1951}.  
Conversely, non-homogeneus metamaterial structures enable the implementation of either positive or negative $\mu$ and $\epsilon$, opening the way to new and exciting design options.  Of course, $\mu$ and $\epsilon$ are now local properties which hold in a given frequency range, and the refraction index has to be re-defined as $n{\,=\,}\pm\sqrt{\epsilon\mu}$ or $n{\,=\,}j\sqrt{\epsilon\mu}$, depending on the combination of signs \cite{Poddar-2014}.  
In a resonator implemented with $\epsilon{\,<\,}0$ and $\mu{\,<\,}0$, hence $n{\,=\,}-\sqrt{\epsilon\mu}$, the noise sources degenerate into magneto-inductive noise, leading to the propagation in the form of forward and backward noise waves \cite{Smith-2002-PRB}.  Any material supporting single propagating mode at a given frequency has a well-defined index $n$ and normalized impedance $z$, whether the material is homogeneous and continuous or not.  By contrast, it is not easy to assign a normalized impedance $z$ to a non-homogeneous material \cite{Syms-2013-PRB}.  Multi-mode random spectral signals crossing a negative-index bandpass filter, hitting simultaneously on the two channels, cannot be rejected due to ambiguity associated with wave impedances (forward $z^{+}$, and backward $z^{-}$).

\begin{figure}\centering
\includegraphics[angle=90, scale=0.9]{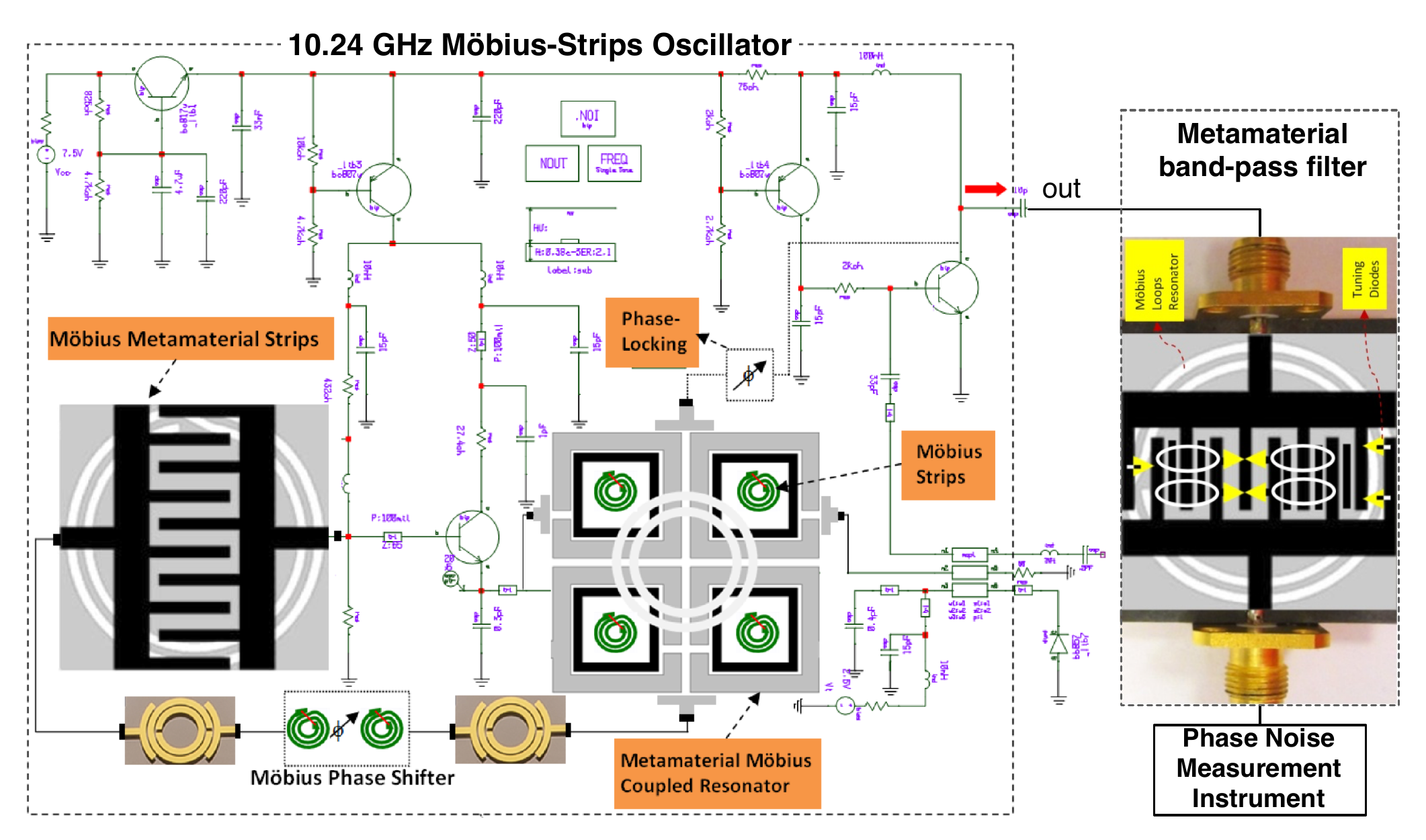}
\caption{Scheme of our 10.24 GHz DRO using M\"{o}bius Coupled Dielectric Resonator.}
\label{fig:Mobius-oscillator}
\end{figure}

We measured the noise of an innovative oscillator, where the resonator is coupled to the electronics through negative-index M\"{obius} structure.  There results a quality factor $Q$ of about 25\,000.  The complete unit (Fig.\,\ref{fig:Mobius-oscillator}) is packaged in a ${\approx}80$ \unit{cm^3} connectorized aluminum case, takes 400 mW DC power (8 V, 50 mA), and delivers $+15$ dBm at 10.24 GHz. 
A \emph{M\"{o}bius metamaterial filter}, shown on the left-hand side of Fig.\,\ref{fig:Mobius-oscillator}, is inserted between the oscillator and the phase-noise measurement instrument.  Four different configurations are tested, which implement the combinations of positive and negative $\epsilon$ and $\mu$ by tuning the varactor diodes.  The insertion loss is 3--5 dB, and the bandwidth of 0.5--1 MHz.  Tuning the diodes, a bandwidth up to 100 MHz can be obtained.  The reference \cite{Poddar-2016-IFCS-Mobius} delivers details about use of metamaterial tunable resonators and filters for VCO applications.
The instrument is a Rohde \& Schwarz FSUP\,26, implementing digital signal processing on the IF signal after I-Q down conversion, and of course cross-spectrum analysis.  Relevant to us, it has a 3 dB coupler as the input power splitter.  Unlike the saturated mixer, the hardware does not phase lock to the input, nor uses a reference at the same frequency.  Common sense suggests that artifacts related to injection locking are virtually impossible.
The measurements were performed at Synergy Microwave Corp.\ in a Faraday cage.  

\begin{figure}\centering
\includegraphics[scale=0.707]{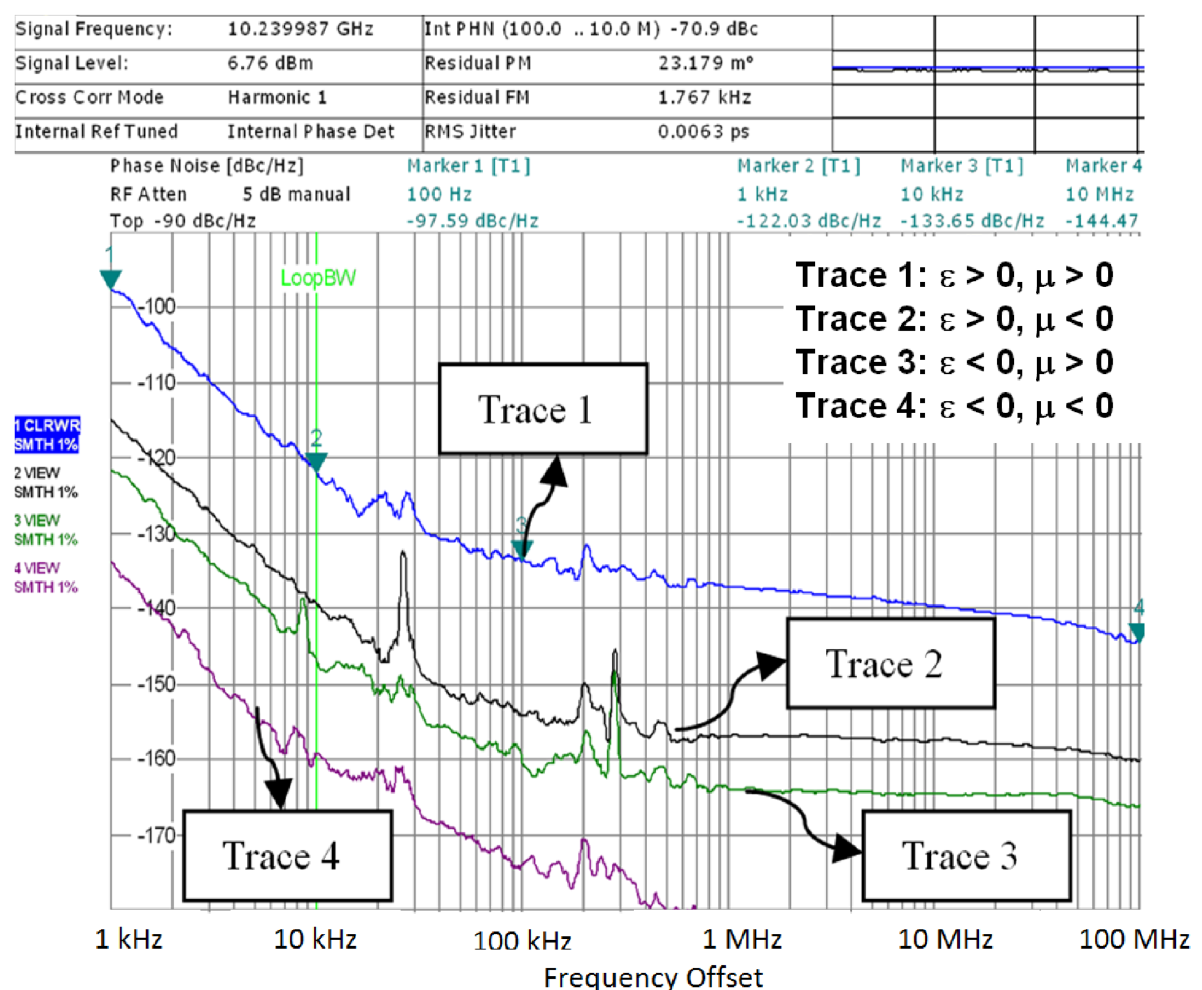}
\caption{Phase noise measured with a metamateral filter between the oscillator under test and the phase-noise measurement instrument.}
\label{fig:xSp-Meta}
\end{figure}

Figure\,\ref{fig:xSp-Meta} shows the phase noise spectra.
Each plot results from averaging on 10\,000 cross spectra.  After observing issues about the convergence of the average, we repeated the measurement several times and we selected the best run. 
Looking at the four traces, we notice a spread of about 40 dB, which is alarming.  Trace\,1, which is the reference to the extent that the filter has $\epsilon{\,>\,}0$ and $\mu{\,>\,}0$, shows the highest phase noise.  Lacking an appropriate model which include the metamaterial properties, the multi-mode spectrum can be misleading or wrong.
The little understanding we have is discussed below.

The negative-index resonator shows distinct values of the impedance $z$ for the waves propagating in forward or backward direction. This happens because the ratio of electric/magnetic field varies periodically throughout the structure.  Metamaterials are both nonlinear and dispersive, which adds complexity to the problem.   
In another occasion \cite{Poddar-2016-IFCS-Validity} we observed that multiple and physically separated tuning diodes connected to negative-index resonator yield a reduction in noise, as compared to the single tuning diode --- which is of course larger in order to provide equivalent range. A possible explanation could be that evanescent-mode EM coupling between the diodes lowers the noise.  
Anyway, a similar phenomenon was observed on `regular,' $n>0$, resonators \cite{Underhill-2016}.

\section{Conclusions}
The thermal energy associated to the dissipation inside the input power splitter introduces a bias $\Delta S_\varphi$ given by \req{eqn:Sphi-cpl-error} for the loss-free directional coupler terminated at one end, and by \req{eqn:Sphi-Y-error} for the Y resistive splitter.

Most often, the equivalent temperature $T_C$ at the oscillator output is significantly higher than the environment temperature $T$\@.  If so, neglecting the bias results in a satisfactory approximation.  However, in the case of high-purity oscillators the result can be wrong.

A radical solution consists of cooling down the input splitter and the dark port, and accounting for the cryostat temperature.  Just cooling the dark port, as often done in radiometry with dedicated low-loss splitters, may not be sufficient here because the splitter's dissipation contributes to noise (see \cite[Fig.\,1 and related text]{Rubiola-2000-RSI} for a lossy path at inhomogeneous temperature). 
Notice that liquid \unit{N_2} is 5.8 dB cooler than room temperature, and it takes liquid He to get 18.5 dB\@.

Moving the input splitter and the dark port to inside the cryostat may be mandatory for the measurement of cryogenic oscillators.  However, cryogeny is hardly compatible with regular RF/microwave instruments, intended for room-temperature oscillators.  In this case, the correction $\Delta S_\varphi$ can be introduced in the readout.  Should the instrument be able to measure the input power at the same time as phase noise, the correction would just be a matter of software update.  
Whether to opt for the simple use of \req{eqn:Sphi-cpl-error} or \req{eqn:Sphi-Y-error}, or to include impedance matching and losses in the model, is a matter of tradeoff between accuracy, complexity, feasibility, and cost.

In challenging measurements, the estimator $\Re\{\left<S_\varphi(f)\right>\}$ provides useful diagnostic power and faster averaging, as compared to $|\left<S_\varphi(f)\right>|$. 

Finally, our experiments warn against trusting any result if a metamaterial behavior is present between the oscillator output and the input of the instrument.  A spread of 40 dB in the measurement of an oscillator has been observed, just changing the nature of the metamaterial filter, and none of the available models helps. 
Such highly erratic behavior appeals for theoretical analysis.

\section*{Acknowledgements}
This work is partially funded by the ANR ``Programme d'Investissement d'Avenir'' (PIA) under the Oscillator\,IMP project and First-TF network, and by grants from the R\'{e}gion Franche Comt\'{e} intended to support the PIA.
We owe gratitude to Prof.\ Michele Elia (retired, Politecnico di Torino, Italy) for inspiring discussions on the mathematical approach.  We thank Dr.\ Rodolphe Boudot (FEMTO-ST Institute) for the mumetal shields and for help; Mr.\ Philippe Abb\'{e} (FEMTO-ST Institute) for his hard and smart work setting up the shielded chamber and the environment stabilization; Dr.\ Ignaz Eisele (EMFT Fraunhofer Institute, Munich, Germany) for developing the prototype of the metamaterial filter.

\def\bibfile#1{/Users/rubiola/CloudStation/Drive/Author/!-Bibliography/#1}
\bibliographystyle{IEEEtran}
\bibliography{\bibfile{ref-short},%
              \bibfile{references},%
              \bibfile{rubiola},%
              Local}

\begin{thebibliography}{10}
\providecommand{\url}[1]{#1}
\csname url@samestyle\endcsname
\providecommand{\newblock}{\relax}
\providecommand{\bibinfo}[2]{#2}
\providecommand{\BIBentrySTDinterwordspacing}{\spaceskip=0pt\relax}
\providecommand{\BIBentryALTinterwordstretchfactor}{4}
\providecommand{\BIBentryALTinterwordspacing}{\spaceskip=\fontdimen2\font plus
\BIBentryALTinterwordstretchfactor\fontdimen3\font minus
  \fontdimen4\font\relax}
\providecommand{\BIBforeignlanguage}[2]{{%
\expandafter\ifx\csname l@#1\endcsname\relax
\typeout{** WARNING: IEEEtran.bst: No hyphenation pattern has been}%
\typeout{** loaded for the language `#1'. Using the pattern for}%
\typeout{** the default language instead.}%
\else
\language=\csname l@#1\endcsname
\fi
#2}}
\providecommand{\BIBdecl}{\relax}
\BIBdecl

\bibitem{Hanbury-Brown-1952}
R.~Hanbury~Brown, R.~C. Jennison, and M.~K. Das~Gupta, ``Apparent angular sizes
  of discrete radio sources,'' \emph{Nature}, vol. 170, no. 4338, pp.
  1061--1063, Dec.~20, 1952.

\bibitem{Vessot-1964}
R.~F.~C. Vessot, R.~F. Mueller, and J.~Vanier, ``A cross-correlation technique
  for measuring the short-term properties of stable oscillators,'' in
  \emph{Proc.\ {IEEE-NASA} Symposium on Short Term Frequency Stability},
  Greenbelt, {MD}, {USA}, Nov.~23-24 1964, pp. 111--118.

\bibitem{Walls-1976}
F.~L. Walls, S.~R. Stain, J.~E. Gray, and D.~J. Glaze, ``Design considerations
  in state-of-the-art signal processing and phase noise measurement systems,''
  in \emph{Proc.\ Freq.\ Control Symp.}, Atlantic City, {NJ}, {USA}, Jun.~2-4
  1976, pp. 269--274.

\bibitem{WWalls-1992}
W.~F. Walls, ``Cross-correlation phase noise measurements,'' in \emph{Proc.\
  Freq.\ Control Symp.}, Hershey, {PA}, May~27-29 1992, pp. 257--261.

\bibitem{Nelson-2014}
C.~W. Nelson, A.~Hati, and D.~A. Howe, ``A collapse of the cross-spectral
  function in phase noise metrology,'' \emph{Rev.\ Sci.\ Instrum.}, vol.~85,
  no.~3, pp. 024\,705 1--7, Mar. 2014.

\bibitem{Paris-2014}
``{E}uropean {W}orkshop on {C}ross-{S}pectrum {P}hase {N}oise {M}easurements,''
  LNE Central Site, Paris, France, Dec.~ 18, 2014.

\bibitem{Denver-2015}
``{C}ross {S}pectrum $\mathscr{L}(f)$ {W}orkshop,'' Denver, CO, USA, Apr.~ 15,
  2015.

\bibitem{Allred-1962}
C.~M. Allred, ``A precision noise spectral density comparator,'' \emph{J. Res.\
  NBS}, vol. 66C, pp. 323--330, Oct.-Dec. 1962.

\bibitem{White-1996-Thermometry}
D.~R. White, R.~Galleano, A.~Actis, H.~Brixy, M.~De~Groot, J.~Dubbeldam, A.~L.
  Reesink, F.~Edler, H.~Sakurai, R.~L. Shepard, and G.~J. C., ``The status of
  {Johnson} noise thermometry,'' \emph{Metrologia}, vol.~33, pp. 325--335,
  1996.

\bibitem{Rubiola-2000-RSI}
E.~Rubiola and V.~Giordano, ``Correlation-based phase noise measurements,''
  \emph{Rev.\ Sci.\ Instrum.}, vol.~71, no.~8, pp. 3085--3091, Aug. 2000.

\bibitem{Rubiola-2002-RSI}
------, ``Advanced interferometric phase and amplitude noise measurements,''
  \emph{Rev.\ Sci.\ Instrum.}, vol.~73, no.~6, pp. 2445--2457, Jun. 2002.

\bibitem{Hati-2016-xSpectrum}
A.~Hati, C.~W. Nelson, and D.~A. Howe, ``Cross-spectrum measurement of
  thermal-noise limited oscillators,'' \emph{Rev.\ Sci.\ Instrum.}, vol.~87,
  pp. 2596--2598, Nov. 2012.

\bibitem{Kock-1946}
W.~E. Kock, ``Metal-lens antennas,'' \emph{Proc.\ IRE}, vol.~34, no.~11, pp.
  828--836, Nov. 1946.

\bibitem{Shelby-2001}
R.~A. Shelby, D.~R. Smith, and S.~S., ``Experimental verification of a negative
  index of refraction,'' \emph{Science}, vol. 292, no. 5514, pp. 77--79, Apr.
  2001.

\bibitem{Pond-2000}
J.~M. Pond, ``M\"{o}bius dual-mode resonators and bandpass filters,''
  \emph{IEEE Trans.\ Microw.\ Theory Tech.}, vol.~48, no.~12, pp. 2465--2471,
  2000.

\bibitem{Poddar-2016-IFCS-Mobius}
A.~K. Poddar, U.~L. Rohde, I.~Eisele, and E.~Rubiola, ``{NIMS} ({N}egative
  {I}ndex {M}\"{o}bius {S}trips): Resonator for next generation electronic
  signal sources,'' in \emph{Proc.\ Int'l Freq.\ Control Symp.}, New Orleans,
  LA, USA, May~9--12 2016, pp. 575--584.

\bibitem{Rohde-2016-MJ1}
U.~L. Rohde and A.~K. Poddar, ``M\"{o}bius metamaterial inspired next
  generation circuits and systems,'' \emph{Microw. J.}, pp. 62--90, May 2016.

\bibitem{Rohde-2016-MJ2}
------, ``M\"{o}bius metamaterial inspired next generation circuits and
  sensors,'' \emph{Microw. J.}, pp. 60--94, Jun. 2016.

\bibitem{Rohde-2016-MJ3}
------, ``M\"{o}bius metamaterial strips: Opportunity, trends, challenges and
  future,'' \emph{Microw. J.}, pp. 62--96, Jul. 2016.

\bibitem{Poddar-2014-habil}
A.~K. Poddar, ``Slow wave resonator based tunable multi-band multi-mode
  injection-locked oscillators,'' Dr.-Ing.-habil Thesis, BTU Cottbus, Cottbus,
  Germany, Apr.~22, 2014.

\bibitem{Rubiola-2010-arXiv-xSpectrum}
E.~Rubiola and F.~Vernotte, ``The cross-spectrum experimental method,''
  ar{X}iv:1004.5539 [physics.ins-det], Apr. 2010.

\bibitem{Keysight:E5500}
\emph{Keysight {E5500} Series, Phase Noise Measurement Solutions}, Keysight
  Technologies, Inc., Paloalto, CA, 2004, document 5989-0851EN, available
  online on the web site http://www.keysight.com.

\bibitem{Rohde-1978}
U.~L. Rohde, ``Mathematical analysis and design of ultra stable low noise 100
  {MHz} crystal oscillator with differential limiter and its possibilities in
  frequency standards,'' in \emph{Proc.\ Freq.\ Control Symp.}, May 1978, pp.
  409--425.

\bibitem{Rohde-1975-ED}
------, ``Crystal oscillator provides low noise,'' \emph{Electronic Design},
  no.~21, p. 11 and 14, Oct.~11, 1975.

\bibitem{Rohde:Synthesizers}
------, \emph{Microwave and Wireless Synthesizers}.\hskip 1em plus 0.5em minus
  0.4em\relax New York: John Wiley \& Sons, 1997.

\bibitem{Giacoletto-1969}
L.~J. Giacoletto, ``Diode and transistor equivalent circuits for transient
  operation,'' \emph{IEEE J. Solid-State Circuits}, vol.~4, no.~2, Feb. 1969.

\bibitem{Gummel-Poon-1970}
H.~K. Gummel and H.~C. Poon, ``An integral charge control model of bipolar
  transistors,'' \emph{Bell Syst.\ Techn.\ J.}, vol.~49, pp. 827--852, May
  1970.

\bibitem{Niu-2001}
G.~Niu, J.~D. Cressler, S.~Zhang, W.~E. Ansley, C.~S. Webster, and D.~L.
  Harame, ``A unified approach to {RF} and microwave noise parameter modeling
  in bipolar transistors,'' \emph{IEEE Trans.\ Electron.\ Dev.}, vol.~48,
  no.~11, pp. 2568--2574, 2001.

\bibitem{Voinigescu-1997}
S.~P. Voinigescu, M.~C. Maliepaard, J.~L. Showell, G.~E. Babcock, D.~Marchesan,
  M.~Schroter, P.~Schvan, and D.~L. Harame, ``A scalable high-frequency noise
  model for bipolar transistors with application to optimal transistor sizing
  for low-noise amplifier design,'' \emph{IEEE J. Solid-State Circuits},
  vol.~32, no.~9, pp. 1430--1439, 1997.

\bibitem{Vasilescu-2005}
G.~Vasilescu, \emph{Electronic Noise and Interfering Signals}.\hskip 1em plus
  0.5em minus 0.4em\relax Berlin, Germany: Springer, 2005.

\bibitem{Driscoll-1995-FCS}
M.~M. Driscoll and R.~W. Weinert, ``Spectral performance of sapphire dielectric
  resonator-controlled oscillators operating in the 80 k to 275 k temperature
  range,'' in \emph{Proc.\ Freq.\ Control Symp.}, San Francisco, CA, USA,
  May~31 - Jun.~2 1995, pp. 401--412.

\bibitem{Callen-Welton-1951}
H.~B. Callen and T.~A. Welton, ``Irreversibility and generalized noise,''
  \emph{Phys.\ Rev.}, vol.~83, no.~1, pp. 34--40, Jul.~1, 1951.

\bibitem{Poddar-2014}
A.~K. Poddar and U.~L. Rohde, ``Metamaterial {M}\"{o}bius strips ({MMS}):
  Application in resonators for oscillators and synthesizers,'' in \emph{Proc.\
  Int'l Freq.\ Control Symp.}, May~19--22, 2014, pp. 1--9.

\bibitem{Smith-2002-PRB}
D.~R. Smith, S.~Schultz, P.~Marko\v{s}, and C.~M. Soukoulis, ``Determination of
  effective permittivity and permeability of metamaterials from reflection and
  transmission coefficients,'' \emph{Phys.\ Rev.\ B}, vol.~65, p. 195104, Apr.
  2002.

\bibitem{Syms-2013-PRB}
R.~R.~A. Syms, O.~Sydoruk, and L.~Solymar, ``Noise in one-dimensional
  metamaterials supporting magnetoinductive lattice waves,'' \emph{Phys.\ Rev.\
  B}, vol.~87, p. 155155, Apr. 2013.

\bibitem{Poddar-2016-IFCS-Validity}
A.~K. Poddar, U.~L. Rohde, E.~Rubiola, and K.-H. Hoffman, ``Validity of
  cross-spectrum pn measurement,'' in \emph{Proc.\ Int'l Freq.\ Control Symp.},
  New Orleans, LA, USA, May~9--12 2016, pp. 12--17.

\bibitem{Underhill-2016}
M.~J. Underhill, ``Oscillator resistor noise optimisation paradigm,'' in
  \emph{Proc.\ Int'l Freq.\ Control Symp.}, New Orleans, LA, USA, May~9--12
  2016, pp. 377--381.

\end{thebibliography}

\end{document}